\documentclass[10pt,twocolumn]{article}%
\usepackage[brazil]{babel}
\usepackage{amsmath,amssymb,amsfonts}
\usepackage{url}
\usepackage{graphicx}
\usepackage{amsmath}
\usepackage{amsfonts}
\usepackage{amssymb}%
\begin{document}

\title{O B\'{o}son de Higgs\\{\small (The Higgs boson)}}
\author{Jean J\'{u}nio Mendes Pimenta$^{^{1}}$, Lucas Francisco Bosso Belussi$^{^{1}}%
$, \'Erica Regina Takano Natti$^{^{2}}$\\Paulo Laerte Natti$^{^{3}}${\thanks{E-mail: plnatti@uel.br}}\\\textit{{\scriptsize $^{^{1}}$Departamento de F\'{\i}sica, Universidade
Estadual de Londrina, Londrina, PR - Brasil}}\\\textit{{\scriptsize $^{^{2}}$Pontif\'{\i}cia Universidade Cat\'olica do
Paran\'a, Campus Londrina, Londrina, PR - Brasil}}\\\textit{{\scriptsize $^{^{3}}$Departamento de Matem\'{a}tica, Universidade
Estadual de Londrina, Londrina, PR - Brasil}}}
\maketitle

\begin{abstract}
O b\'oson de Higgs foi predito em 1964 pelo f\'{\i}sico brit\^anico Peter
Higgs. O Higgs representa a chave para explicar a origem da massa das outras
part\'{\i}culas elementares da Natureza. Entretanto, somente com a entrada em
funcionamento do LHC, em 2008, houve condi\c{c}\~oes tecnol\'ogicas para a
procura pelo b\'oson de Higgs. Recentemente, num grande esfor\c{c}o
internacional realizado no CERN, por meio dos experimentos ATLAS e CMS, foi
observada uma nova part\'{\i}cula bos\^onica na regi\~ao de 125 GeV. Neste
artigo, por meio dos conhecidos mecanismos de quebra de simetria que ocorrem
na teoria BCS da supercondutividade e na teoria do emparelhamento nuclear,
discute-se o mecanismo de Higgs no Modelo Padr\~ao. Enfim, apresentamos a
situa\c{c}\~ao atual da procura pelo b\'oson de Higgs e as teorias
alternativas e extens\~{o}es do Modelo Padr\~ao para a f\'{\i}sica de
part\'{\i}culas elementares.

\noindent\textbf{Palavras-chave:} b\'{o}son de Higgs, teoria BCS,
emparelhamento nuclear, mecanismo de Higgs, Modelo Padr\~{a}o

\vskip 1.0cm

The Higgs boson was predicted in 1964 by British physicist Peter Higgs. The
Higgs is the key to explain the origin of the mass of other elementary
particles of Nature. However, only with the coming into operation of the LHC,
in 2008, there were technological conditions to search for the Higgs boson.
Recently, a major international effort conducted at CERN, by means of ATLAS
and CMS experiments, has enabled the observation of a new bosonic particle in
the region of 125 GeVs. In this paper, by means of known mechanisms of
symmetry breaking that occur in the BCS theory of superconductivity and in the
theory of nuclear pairing, we discuss the Higgs mechanism in the Standard
Model. Finally, we present the current state of research looking for the Higgs
boson and the alternative theories and extensions of the Standard Model for
the elementary particle physics.


\end{abstract}

\section{Introdu\c{c}\~ao}

\indent Em 4 de julho de 2012, cientistas do CERN (Organiza\c{c}\~{a}o
Europ\'{e}ia para a Investiga\c{c}\~{a}o Nuclear) anunciaram que, ao fim de 50
anos de investiga\c{c}\~{a}o, foi descoberta uma nova part\'{\i}cula que pode
ser o b\'{o}son de Higgs \cite{noticia}. O an\'{u}ncio da descoberta do Higgs
foi feito com grande cautela. Dois grupos de pesquisadores, das
experi\^{e}ncias ATLAS e CMS, anunciaram a descoberta de uma part\'{\i}cula,
com massa entre 125 e 126 GeV, que possui propriedades semelhantes \`{a}quelas
previstas para o Higgs do Modelo Padr\~{a}o. Contudo, ainda \'{e}
necess\'{a}rio verificar se essa part\'{\i}cula possui todas as propriedades
do Higgs, devendo-se ent\~{a}o acumular mais dados experimentais. Por exemplo,
os dados experimentais mostram que devido ao fato da part\'{\i}cula ter
deca\'{\i}do em pares de f\'{o}tons ela pode ter spin zero ou dois,
consistente com o Higgs do Modelo Padr\~{a}o que tem spin zero. Por outro
lado, de acordo com o Modelo Padr\~{a}o, o Higgs deveria decair com uma
determinada frequ\^{e}ncia em pares de f\'{o}tons, mas os dados preliminares
experimentais acusam um excesso no n\'{\i}vel deste tipo de decaimento. Outro
dado experimental inconsistente com o Higgs do Modelo Padr\~{a}o \'{e} uma
car\^{e}ncia no decaimento em part\'{\i}culas taus. Estes resultados
indicariam que o b\'{o}son encontrado n\~{a}o seria o Higgs do Modelo
Padr\~{a}o. Assim, somente com os dados experimentais que ser\~{a}o acumulados
nos pr\'{o}ximos meses ou anos ser\'{a} poss\'{\i}vel ter uma maior clareza do
que foi observado em 4 de julho. Estes novos dados experimentais tamb\'{e}m
permitir\~{a}o testar os modelos alternativos e as extens\~{o}es do Modelo
Padr\~{a}o, tais como os Modelos Supersim\'{e}tricos, os Modelos de
Dimens\~{o}es Extras, os Modelos Technicolor, entre outros. Nos pr\'{o}ximos
meses ou anos certamente veremos grandes mudan\c{c}as no entendimento a
respeito das part\'{\i}culas fundamentais da natureza.

Comecemos explicando quem \'{e} o Higgs do Modelo Padr\~{a}o. O Higgs \'{e}
uma part\'{\i}cula elementar bos\^{o}nica (part\'{\i}cula com spin inteiro),
proposta para validar o Modelo Padr\~{a}o, que seria respons\'{a}vel pela
origem da massa de todas as outras part\'{\i}culas elementares. Sobre o
conceito de massa na teoria da Relatividade veja as refer\^encias
\cite{Castellani2001,Goto2010}. Na Mec\^anica Qu\^antica toda part\'{\i}cula
elementar \'e associada a um campo. Assim, quando o campo de Higgs,
que permeia todo o universo, recebe energia suficiente, ele cria uma
part\'{\i}cula, o Higgs, que \'e uma excita\c{c}\~{a}o do campo de Higgs.
Por outro lado, quando a part\'{\i}cula de Higgs interage com outras
part\'{\i}culas elementares (el\'etrons, quarks,..), ela transfere energia,
na forma de massa, do campo de Higgs para a part\'{\i}cula elementar.
Lembre-se que massa \'e uma forma de energia. Portanto, dependendo da
intensidade da intera\c{c}\~{a}o do Higgs com uma part\'{\i}cula
elementar, o campo de Higgs determina a "quantidade" massa desta
part\'{\i}cula. Analogamente, sabemos que um el\'etron ao interagir
com um f\'oton na presen\c{c}{a} de um campo eletromagn\'etico ganha
(ou perde) energia, na forma cin\'etica.

Fora da
comunidade cient\'{\i}fica o b\'{o}son de Higgs \'{e} mais conhecido como a
part\'{\i}cula de Deus (\textit{God particle}), nome popularizado pelo
f\'{\i}sico Leon Lederman em seu livro de divulga\c{c}\~{a}o cient\'{\i}fica,
The God Particle: If the Universe Is the Answer, What Is the Question?
\cite{Lederman}. Nesse livro Lederman fez uma analogia entre a propriedade do
Higgs de dar diferentes massas \`{a}s diferentes part\'{\i}culas elementares e
a hist\'{o}ria b\'{\i}blica da Torre de Babel, em que Deus, num acesso de
f\'{u}ria, fez com que todos falassem l\'{\i}nguas diferentes. A comunidade
cient\'{\i}fica prefere cham\'{a}-lo de b\'{o}son de Higgs, j\'{a} que
\textit{God particle} \'{e} inapropriado devido \`{a}s conota\c{c}\~{o}es
religiosas que o nome possa sugerir.

Do ponto de vista hist\'{o}rico, Sheldon Lee Glashow, em 1961, descreveu uma
teoria que unificava a intera\c{c}\~{a}o eletromagn\'{e}tica, que atua sobre
part\'{\i}culas carregadas, e a intera\c{c}\~{a}o fraca, que atua sobre
quarks, el\'{e}trons e neutrinos \cite{Glashow1961}. Experimentalmente, devido
ao curto alcance da intera\c{c}\~{a}o fraca, as part\'{\i}culas mediadoras da
teoria (b\'{o}sons vetoriais) deveriam ser muito massivas. Contudo, para
garantir a renormalizabilidade do Modelo de Glashow, as part\'{\i}culas
mediadoras das intera\c{c}\~{o}es eletromagn\'{e}tica e fraca n\~{a}o deveriam
ser massivas. Para resolver este problema, Robert Brout e Fran\c{c}ois Englert
\cite{Brout-Englert:1964}, Gerald Guralnik, Carl Richard Hagen e Tom Kibble
\cite{Guralink-Hagen-Kibble:1964} e Peter Ware Higgs \cite{PeterHiggs:1964},
sugeriram, independentemente, que os b\'{o}sons vetoriais $W^{+}$, $W^{-}$ e
$Z_{0}$ adquirissem massa devido a um campo bos\^{o}nico, o campo de Higgs. A
proposta do campo de Higgs era apoiada em teorias j\'{a} conhecidas neste
per\'{\i}odo, que descreviam gera\c{c}\~{a}o de massa por meio de um
fen\^{o}meno de quebra de simetria, ou seja, a teoria BCS para a
supercondutividade a baixas temperaturas
\cite{Bardeen-Cooper-Schrieffer:1957,kittel} e a teoria do emparelhamento
nuclear no modelo da gota l\'{\i}quida \cite{Weis:1935,almeida:tauhata}.

Nesse contexto, entre 1967 e 1968, Steven Weinberg \cite{Weinberg:1967} and
Mohammad Abdus Salam \cite{Salam:1968}, incorporando o mecanismo de
gera\c{c}\~{a}o de massa de Higgs no Modelo Eletrofraco de Glashow
\cite{Glashow1961}, propuseram a forma atual do Modelo Eletrofraco. Nessa
\'{u}ltima forma, os b\'{o}son vetoriais $W^{+}$, $W^{-}$ e $Z_{0}$ e os
f\'{e}rmions, quarks e l\'{e}ptons, adquirem massa por meio de uma
transi\c{c}\~{a}o de fase dependente da temperatura do universo. Assim, no
in\'{\i}cio do universo, os b\'{o}sons da for\c{c}a eletrofraca n\~{a}o eram
massivos. Mas, \`{a} medida que a temperatura do universo diminuiu, foi
atingida uma temperatura cr\'{\i}tica, correspondente a uma transi\c{c}\~{a}o
de fase, em que foi gerado um campo de for\c{c}a, chamado campo de Higgs.
Enfim, no Modelo Eletrofraco \'{e} a intera\c{c}\~{a}o do campo de Higgs com
os b\'{o}sons vetoriais que gera a massa para esses \'{u}ltimos.

J\'{a} no in\'{\i}cio dos anos setenta foi desenvolvido o Modelo Padr\~{a}o da
f\'{\i}sica de part\'{\i}culas elementares. O Modelo Padr\~{a}o \'{e} uma
teoria que pretende descrever de forma unificada as for\c{c}as forte, fraca e
eletromagn\'{e}tica, bem como todas as intera\c{c}\~{o}es entre as
part\'{\i}culas fundamentais que constituem a mat\'{e}ria, exceto a
intera\c{c}\~{a}o gravitacional. Esta teoria qu\^{a}ntica de campos \'{e}
consistente com a mec\^{a}nica qu\^{a}ntica e com a relatividade especial. No
Modelo Padr\~{a}o os l\'{e}ptons e quarks s\~{a}o considerados constituintes
fundamentais da mat\'{e}ria e a intera\c{c}\~{a}o entre eles ocorre por meio
da troca de b\'{o}sons (part\'{\i}culas mediadoras). Os b\'{o}sons do Modelo
Padr\~{a}o s\~{a}o o f\'{o}ton (intera\c{c}\~{a}o eletromagn\'{e}tica), os
gl\'{u}ons (intera\c{c}\~{a}o forte) e as part\'{\i}culas $W^{+}$, $W^{-}$ e
$Z_{0}$ (intera\c{c}\~{a}o fraca). O Modelo Padr\~{a}o tamb\'{e}m prev\^{e} a
exist\^{e}ncia do chamado campo de Higgs, que permeia todo o Universo dando
massa a todas as part\'{\i}culas que interagem com ele, inclusive aos
b\'{o}sons vetoriais, como predito no Modelo Eletrofraco. De acordo com o
Modelo Padr\~{a}o, quanto maior for a intera\c{c}\~{a}o de uma part\'{\i}cula
com o campo de Higgs, maior ser\'{a} a massa desta part\'{\i}cula, enquanto as
part\'{\i}culas que n\~{a}o interagem com o campo de Higgs t\^{e}m massa nula
\cite{:Moreira:2009:}.

V\'{a}rias experi\^{e}ncias conduzidas no in\'{\i}cio dos anos setenta
confirmaram as previs\~{o}es do Modelo Eletrofraco e do Modelo Padr\~{a}o. Em
particular, em 1973, verificou-se experimentalmente no CERN a ocorr\^{e}ncia
da corrente fraca neutra devido \`{a} troca de b\'{o}sons $Z_{0}$
\cite{Hasert:1974,fortes:2007}. Estes resultados experimentais fizeram com
que, em 1979, Salam, Weinberg e Glashow, fossem agraciados com o Pr\^{e}mio
Nobel por unificar as for\c{c}as fraca e eletromagn\'{e}tica por meio do
Modelo Eletrofraco. Ainda neste per\'{\i}odo, entre 1973-1974, citamos
tamb\'{e}m a descoberta do quark $c$ \cite{quarkC:1974}, previsto em 1970 por
Sheldon Lee Glashow, John Iliopoulos e Luciano Maiani \cite{Iliopoulos:1970}.
At\'{e} o momento j\'{a} foram observados seis tipos de quarks, tamb\'{e}m
chamados sabores, que s\~{a}o os quarks $u$ (up), $d$ (down), $s$ (strange),
$c$ (charm), $b$ (bottom) e $t$ (top) \cite{oldoni.natti:2007}. Finalmente
citamos os experimentos UA1 e UA2, conduzidos no CERN, em 1983, por mais de
100 f\'{\i}sicos liderados por Carlo Rubbia \cite{rubia1,rubia2}, que
verificaram a exist\^{e}ncia e mediram as massas dos b\'{o}sons vetoriais
$W^{+}$, $W^{-}$ e $Z_{0}$ \cite{Amsler20081}. Carlo Rubbia e Simon van der
Meer receberam o Pr\^{e}mio Nobel de 1984 por estas observa\c{c}\~{o}es
experimentais que consolidaram o Modelo Padr\~{a}o das Part\'{\i}culas Elementares.

Apesar do Modelo Padr\~{a}o ser bem sucedido para explicar os processos
fundamentais da natureza, h\'{a} problemas, como a necessidade da
detec\c{c}\~{a}o do b\'{o}son de Higgs. Recentemente, os experimentos ATLAS
\cite{ATLAS2012,ATLAS2012v2} e CMS \cite{CMS2012,CMS2012v2}
acabaram de apresentar independentemente
uma atualiza\c{c}\~{a}o da pesquisa sobre o Higgs. Nesses dois experimentos
foi observada a exist\^{e}ncia de uma nova resson\^{a}ncia perto de 125 GeV,
que foi estabelecida com grande precis\~{a}o. V\'{a}rias quest\~{o}es, no
entanto, permanecem em aberto. Qual a natureza precisa desta resson\^{a}ncia?
A resson\^{a}ncia \'{e} um b\'{o}son de Higgs? Ela \'{e} o b\'{o}son de Higgs
do Modelo Padr\~{a}o? Caso contr\'{a}rio, qual \'{e} a nova f\'{\i}sica que
est\'{a} sendo favorecida ou desfavorecida pelos dados sobre o Higgs? Os
pr\'{o}ximos meses ser\~{a}o dedicados a responder quest\~{o}es desse tipo
\cite{ReportHiggs2012}-\cite{ReportHiggs2013CDF}.

O objetivo deste artigo \'{e} apresentar, de forma did\'{a}tica, o Modelo
Padr\~{a}o e o Mecanismo de Higgs, al\'{e}m de discutir a situa\c{c}\~{a}o
atual da procura pelo b\'{o}son de Higgs, mencionando as teorias alternativas
e extens\~{o}es do Modelo Padr\~{a}o para a f\'{\i}sica das part\'{\i}culas
elementares. Na Se\c{c}\~{a}o 2 descreve-se a teoria BCS da supercondutividade
a baixas temperaturas \cite{kittel}, proposta em 1957, com \^{e}nfase na
energia de \textit{gap}. Na Se\c{c}\~{a}o 3 apresenta-se o Modelo da Gota
L\'{\i}quida para a massa nuclear \cite{wong}, de 1935, discutindo-se em
particular o termo de emparelhamento nuclear. Nessas teorias s\~{a}o
utilizados mecanismos de emparelhamento para descrever fen\^{o}menos de
gera\c{c}\~{a}o de massa observados experimentalmente. Na segunda parte do
artigo, na Se\c{c}\~{a}o 4, apresenta-se o Modelo Padr\~{a}o e sua
lagrangiana. Os v\'{a}rios setores da lagrangiana do Modelo Padr\~{a}o s\~{a}o
descritos, inclusive o setor do Higgs, em que h\'{a} um mecanismo de quebra de
simetria an\'{a}logo ao que foi descrito nas teorias BCS e de emparelhamento
nuclear. O leitor n\~{a}o deve se preocupar com o formalismo carregado e a
nota\c{c}\~{a}o da teoria de campos. Na Se\c{c}\~{a}o 5 descreve-se a procura
experimental pelo b\'{o}son de Higgs e os resultados obtidos at\'{e} o
momento. Enfim, na Se\c{c}\~{a}o 6, apresentam-se alguns problemas do Modelo
Padr\~{a}o e as principais teorias alternativas e extens\~{o}es do Modelo
Padr\~{a}o para a f\'{\i}sica das part\'{\i}culas elementares.

\section{Supercondutividade a baixas temperatura}

\noindent Nesta se\c{c}\~{a}o descreve-se a teoria BCS da supercondutividade a
baixas temperaturas \cite{kittel}. Discute-se em particular o mecanismo de
quebra de simetria com gera\c{c}\~{a}o de massa e a express\~{a}o da energia
de \textit{gap}.

\subsection{Teoria BCS}

\begin{figure}[h]
\centering
\includegraphics[scale=0.3]{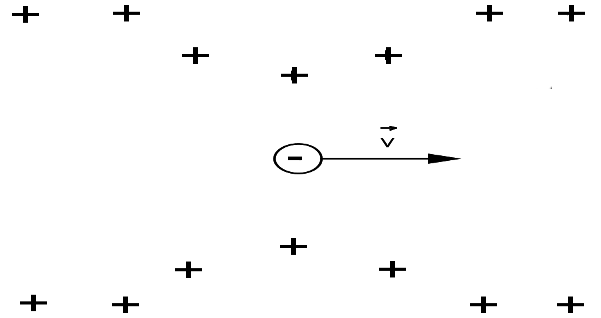}\caption{Esquema pict\'{o}rico da
concentra\c{c}\~ao de \'{\i}ons positivos devido \`a passagem de um
el\'{e}tron.}%
\label{fonon}%
\end{figure}

\noindent John \textbf{B}ardeen, Leon Neil \textbf{C}ooper e John Robert
\textbf{S}chrieffer, em 1957, formularam uma teoria para descrever o
comportamento de materiais que apresentavam o fen\^{o}meno de
supercondutividade a baixas temperaturas. Tal teoria \'{e} conhecida
atualmente como teoria BCS, em homenagem a esses pesquisadores. A seguir,
passamos a descrever a teoria BCS. Sabe-se que \`{a} temperatura ambiente os
materiais possuem el\'{e}trons de condu\c{c}\~{a}o, ou seja, el\'{e}trons
n\~{a}o localizados em um \'{a}tomo particular, os quais est\~{a}o livres para
se deslocarem por todo o material \cite{kittel}. Esta cole\c{c}\~{a}o de
el\'{e}trons livres recebe o nome de Mar de Fermi. Devido \`{a}s suas cargas
negativas, os el\'{e}trons repelem-se entre si, mas tamb\'{e}m atraem os
\'{\i}ons positivos da estrutura do material. Consequentemente, a concentra\c{c}\~{a}o
de \'{\i}ons positivos em uma determinada regi\~{a}o do material aumenta, como
esquematiza a figura \ref{fonon}.

Pelo fato destes \'{\i}ons positivos estarem presos \`{a} rede cristalina, uma
for\c{c}a do tipo restauradora recomp\~oe a estrutura da rede ap\'os a
passagem do el\'etron. Efetivamente, a deforma\c{c}\~ao na regi\~ao de
concentra\c{c}\~ao de carga positiva se propaga como uma onda de
vibra\c{c}\~ao da rede. Ao modo de vibra\c{c}\~ao desta propaga\c{c}\~ao,
d\'{a}-se o nome de f\^{o}non \cite{deforma-rede}.

\begin{figure}[h]
\centering
\includegraphics[scale=0.5]{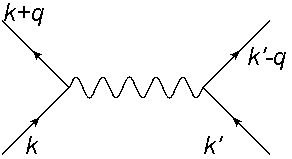}\caption{Esquema da intera\c{c}\~ao
el\'{e}tron-f\^{o}non-el\'{e}tron}%
\label{eletrfoneletr}%
\end{figure}

Nesta interpreta\c{c}\~{a}o da teoria BCS, o el\'{e}tron fornece um momento
para a estrutura do material. Em outras palavras, tudo se passa como se o
el\'{e}tron tivesse emitido um f\^{o}non. Simultaneamente, um segundo
el\'{e}tron, atra\'{\i}do pela concentra\c{c}\~{a}o de cargas positivas, pode
absorver este momento transportado pela onda, ou ainda, conforme a teoria BCS,
este segundo el\'{e}tron pode absorver o f\^{o}non. Neste contexto, o
f\^{o}non induz uma intera\c{c}\~{a}o atrativa el\'{e}tron-el\'{e}tron,
esquematizada na figura \ref{eletrfoneletr}. Note que na figura
\ref{eletrfoneletr} os el\'{e}trons com momentos $\mathbf{k}$ e
$\mathbf{k^{\prime}}$ passam a ter momentos $\mathbf{k}+\mathbf{q}$ e
$\mathbf{k^{\prime}}-\mathbf{q}$, devido \`{a} intera\c{c}\~{a}o com o
f\^{o}non \cite{Dixon:Slac}.

Na teoria BCS h\'{a} uma temperatura cr\'{\i}tica $T_{C}$, que depende do
material, abaixo da qual a atra\c{c}\~{a}o induzida pelo f\^{o}non torna-se
maior do que a repuls\~{a}o coulombiana entre os el\'{e}trons. Assim, para
temperaturas $T<T_{C}$ , ocorre a forma\c{c}\~{a}o de um par de el\'{e}trons
interagentes, ligados pelo f\^{o}non, denominado par de Cooper. Note que o
sistema ligado el\'{e}tron-f\^{o}non-el\'{e}tron (par de Cooper) tem os
el\'{e}trons emparelhados, ou seja, os el\'{e}trons t\^{e}m spins opostos
($\mathbf{k},\uparrow;\mathbf{k^{\prime}},\downarrow$), j\'{a} que esta \'{e}
a configura\c{c}\~{a}o de menor energia. Enfim, o par de Cooper possui spin
zero, inteiro, e, portanto, n\~{a}o obedece ao princ\'{\i}pio de exclus\~{a}o
de Pauli \cite{Piza}. Na teoria BCS, como o par de Cooper \'e um b\'{o}son de spin
zero, ele obedece \`{a} estat\'{\i}stica de Bose-Einstein \cite{Piza}, de modo
que se propaga sem resist\^{e}ncia no material. Dessa forma, o par de Cooper
da teoria BCS explica a supercondutividade de materiais met\'{a}licos a baixas
temperaturas \cite{Costa}.

Na sequ\^{e}ncia deduziremos a energia de \textit{gap} que surge devido \`a
mudan\c{c}a de fase que ocorre em sistema supercondutores \`a temperatura
cr\'{\i}tica $T_{C}$. Discutiremos, tamb\'em, neste contexto, o fen\^omeno de
quebra de simetria com gera\c{c}\~ao de massa.

\subsection{{Energia de \textit{gap}}}

\noindent A energia de gap determina a diferen\c{c}a energ\'{e}tica entre o
estado supercondutor e o estado normal de um material. Materiais na fase
supercondutora t\^{e}m como part\'{\i}culas livres os Pares de Cooper. Para
calcular a energia de um par de Cooper consideremos as coordenadas do centro
de massa deste,%

\begin{equation}
\mathbf{R}=\frac{1}{2}(\mathbf{r_{1}}+\mathbf{r_{2}})\qquad\mathbf{r}%
=\mathbf{r_{1}}-\mathbf{r_{2}}%
\end{equation}%
\begin{equation}
\mathbf{K}=\mathbf{k_{1}}+\mathbf{k_{2}}\qquad\mathbf{k}=\frac{1}%
{2}(\mathbf{k_{1}}+\mathbf{k_{2}})
\end{equation}
\noindent em que $\mathbf{r_{1}}$ e $\mathbf{r_{2}}$ s\~{a}o vetores de
posi\c{c}\~{a}o dos el\'{e}trons e $\mathbf{k_{1}}$ e $\mathbf{k_{2}}$ s\~{a}o
os vetores de momento. Nessas coordenadas escrevemos a equa\c{c}\~{a}o de
Schr\"{o}dinger desse sistema interagente,%

\begin{equation}
(H_{0} + H_{1})\;{\chi}(\mathbf{r}) = \varepsilon\;{\chi}(\mathbf{r}),
\label{hamilt}%
\end{equation}

\noindent em que $H_{0}$ \'{e} a energia cin\'{e}tica e $H_{1}$ \'{e} a
energia potencial, representando uma pertuba\c{c}\~{a}o correspondente \`{a}
intera\c{c}\~{a}o nos pares de Cooper. Enfim, ${\chi}(\mathbf{r})$ e
$\varepsilon$ s\~{a}o respectivamente os autovetores e os autovalores da
equa\c{c}\~{a}o de Schr\"{o}dinger (\ref{hamilt}).

Decompondo os autovetores na base de ondas planas%

\begin{equation}
{\chi}(\mathbf{r) = \sum_{k^{\prime}} G_{k^{\prime}}\;e^{i(k^{\prime} . r)},}
\label{decomp}%
\end{equation}

\noindent e substituindo (\ref{decomp}) em (\ref{hamilt}), obtemos%

\begin{equation}
(E_{\mathbf{k}} - \varepsilon)\;G_{\mathbf{k}} + \sum_{\mathbf{k^{\prime}}}
G_{\mathbf{k^{\prime}}}\;<\mathbf{k}|H_{1}|\mathbf{k^{\prime}}> = 0.
\label{eqeq}%
\end{equation}

\noindent Tomando o limite do cont\'{\i}nuo na equa\c{c}\~{a}o (\ref{eqeq}) e
sendo $N_{f}$ o n\'{u}mero total de estados poss\'{\i}veis de dois
el\'{e}trons (pares de Cooper) com momentos entre $\mathbf{k}$ e
$\mathbf{k}+d\mathbf{k}$, temos%

\begin{equation}
(E - \varepsilon)\;G_{\mathbf{k}} + \displaystyle\int G_{\mathbf{k^{\prime}}%
}\;<\mathbf{k}|H_{1}|\mathbf{k^{\prime}}>\;N_{f} \;d\mathbf{k^{\prime}} = 0.
\label{eqgap}%
\end{equation}

A matriz de Bardeen, Cooper, e Schrieffer, $V_{\mathbf{k},\mathbf{k^{\prime}}%
}=<\mathbf{k}|H_{1}|\mathbf{k^{\prime}}>$, indica a import\^{a}ncia da
intera\c{c}\~{a}o atrativa entre os el\'etrons devido aos f\^{o}nons
em rela\c{c}\~{a}o \`{a} for\c{c}a repulsiva de Coulomb. De forma
expl\'{\i}cita, temos%

\begin{align}
V_{\mathbf{k},\mathbf{k^{\prime}}}  &  = <-\mathbf{k}\uparrow,\mathbf{k}%
\downarrow|H_{1} |-\mathbf{k^{\prime}}\downarrow, \mathbf{k^{\prime}}%
\uparrow>\nonumber\\
&  + <\mathbf{k}\uparrow,-\mathbf{k}\downarrow|H_{1} |\mathbf{k^{\prime}%
}\downarrow, -\mathbf{k^{\prime}}\uparrow>.
\end{align}

Na teoria BCS \cite{Bardeen-Cooper-Schrieffer:1957}, desprezando o efeito de
anisotropia na rede do material, podemos escrever $V_{kk^{\prime}}=-V$, que
deve ser uma constante para pares de Cooper com energia na regi\~{a}o
compreendida entre $-\hbar{\omega}_{D}$ e $\hbar{\omega}_{D}$, em que
${\omega}_{D}$ \'{e} a frequ\^{e}ncia de corte dos f\^{o}nons. Reescrevendo a
equa\c{c}\~{a}o (\ref{eqgap}) na representa\c{c}\~{a}o de energia, temos%

\begin{equation}
(E-\varepsilon)g(E)=V\int_{2{\varepsilon}_{f}}^{2{\varepsilon}_{m}}%
g(E^{\prime})\;N_{f}\;dE^{\prime}, \label{eqgap21}%
\end{equation}

\noindent com ${\varepsilon}_{m}={\varepsilon}_{f}+\hbar{\omega}_{D}$, em que
${\varepsilon}_{f}$ \'{e} a energia na superf\'{\i}cie de Fermi. De
(\ref{eqgap21}) temos $(E-\varepsilon)g(E)=V\;C$, em que $C$ \'{e} uma
constante. Integrando $g(E)=V\;C/(E-\varepsilon)$ e substituindo o resultado
em (\ref{eqgap21}), temos%

\begin{equation}
1=\int_{2{\varepsilon}_{f}}^{2{\varepsilon}_{m}}g(E^{\prime})dE^{\prime}%
=N_{f}\;V\int_{2{\varepsilon}_{f}}^{2{\varepsilon}_{m}}\frac{1}{(E^{\prime
}-\varepsilon)}dE^{\prime}. \label{eqgap22}%
\end{equation}

\noindent Em (\ref{eqgap22}), $\varepsilon$ \'{e} a energia do estado
supercondutor (par de Cooper), ou seja, duas vezes a energia de Fermi menos a
energia de liga\c{c}\~{a}o do par, ou ainda,%

\begin{equation}
\varepsilon=2{\varepsilon}_{f}-\Delta. \label{eqgap23}%
\end{equation}

\noindent Resolvendo a integral (\ref{eqgap22}), utilizando (\ref{eqgap23}),
obtemos a energia de liga\c{c}\~ao $\Delta$ do par de Cooper,%

\begin{align}
1  &  = N_{f} V \ln\left(  \frac{2{\varepsilon}_{m} - 2{\varepsilon}_{f} +
\Delta}{\Delta} \right) \nonumber\\
\Delta &  = \frac{2\hbar{\omega}_{D}}{e^{(\frac{1}{N_{f} V})} - 1}.
\end{align}

Note que a diferen\c{c}a de massa entre dois el\'{e}trons livres e o par de
Cooper \'{e} a energia de liga\c{c}\~{a}o $\Delta$. Este fen\^{o}meno descrito
pelo mecanismo BCS \'{e} denominado quebra de simetria com gera\c{c}\~{a}o de
massa, pois, ao quebrar o par de Cooper com o aumento da temperatura, a
simetria de emparelhamento tamb\'{e}m \'{e} quebrada, e consequentemente
ocorre uma "gera\c{c}\~{a}o" de massa na quantidade $\Delta$. Em outras
palavras, a massa deste sistema \'{e} modificada conforme a simetria criada
pelo emparelhamento dos el\'{e}trons, no estado supercondutor, \'{e} desfeita
(quebrada). Enfim, note que \'{e} o "surgimento" do f\^{o}non que gera a
intera\c{c}\~{a}o entre os el\'{e}trons livres, ou seja, o f\^{o}non \'{e} a
part\'{\i}cula mediadora da intera\c{c}\~{a}o BCS. Veremos adiante que o
b\'{o}son de Higgs foi proposto justamente com este intuito, ou seja, de
mediar uma intera\c{c}\~{a}o com consequente gera\c{c}\~{a}o de massa.

Finalmente salientamos que a supercondutividade observada a altas
temperaturas, $T_{C}>-200^{0}C$, a partir de 1986
\cite{Bednorz-Muller:1986,Wu-Ashburn-Torng},
n\~{a}o \'{e} explicada atrav\'{e}s de um mecanismo
tipo BCS \cite{Tanaka:2001}. Para explicar o comportamento supercondutor a
altas temperaturas algumas teorias foram
desenvolvidas, como por exemplo: a do bipolaron \cite{bipolaron} devido a
Sir Nevill Francis Mott, a da liga\c{c}\~{a}o de val\^encia ressonante
\cite{ressonante} devido a Philip Warren Anderson, a da din\^amica de v\'{o}rtices,
descrita em termos da equa\c{c}\~{a}o de Ginzburg-Landau \cite{Zhou:2001},
entre outras propostas \cite{Costa}. Atualmente j\'{a} s\~{a}o
conhecidas cer\^{a}micas supercondutoras a
temperaturas em torno de $138K=-135^{0}C$ \cite{Dai:1995}.

\section{Emparelhamento nuclear}

\noindent Outro sistema f\'{\i}sico que apresenta um fen\^{o}meno de
emparelhamento, com consequente mecanismo de gera\c{c}\~ao de massa, \'{e} o
nucleo at\^{o}mico. A massa de um sistema n\~ao ligado de $Z$ pr\'{o}tons e
$A-Z$ n\^{e}utrons \'{e} a soma das massas dos constituintes. Sendo $m_{p}$ a
massa do pr\'{o}ton e $m_{n}$ a massa do n\^{e}utron, a massa deste sistema
n\~ao ligado \'{e}
\begin{equation}
M = Zm_{p} + (A-Z)m_{n}. \label{naoligado}%
\end{equation}

Por outro lado, um n\'{u}cleo at\^{o}mico $_{_{Z}}^{^{A}}X$ \'{e} um sistema
de $Z$ pr\'{o}tons e $A-Z$ n\^{e}utrons ligados. Os n\'{u}cleons s\~{a}o
mantidos juntos devido \`{a} for\c{c}a nuclear forte. Para separar os
n\'{u}cleons devemos realizar trabalho, fornecer energia para "romper" a
liga\c{c}\~{a}o entre os componentes nucleares. Da equival\^{e}ncia entre
massa e energia da Relatividade Restrita \cite{Russel}, a massa $M_{n}$ de um
n\'{u}cleo at\^{o}mico com $Z$ pr\'{o}tons e $A-Z$ n\^{e}utrons e a massa de
um sistema n\~{a}o ligado de $Z$ pr\'{o}tons e $A-Z$ n\^{e}utrons obedecem a equa\c{c}\~{a}o%

\begin{equation}
M_{n} (A,Z)c^{2} + B(A,Z)= Zm_{p} c^{2} + (A-Z)m_{n} c^{2},
\label{massanuclear}%
\end{equation}

\noindent ou ainda,%

\begin{equation}
B(A,Z) = [Zm_{p} + (A-Z)m_{n} - M_{n} (A,Z)]c^{2}, \label{Baz}%
\end{equation}

\noindent em que $B(A,Z)$ \'{e} a energia necess\'{a}ria para separar o
n\'{u}cleo em todos os seus n\'{u}cleons, em outras palavras, \'e a energia de
liga\c{c}\~ao do n\'ucleo at\^omico. A energia $B(A,Z)$, definida pela
Equa\c{c}\~ao (\ref{Baz}), \'{e} sempre positiva. Observe a equival\^encia
entre as equa\c{c}\~oes (\ref{eqgap23}) e (\ref{Baz}).

Experimentalmente
verifica-se que a energia de liga\c{c}\~ao por n\'{u}cleon, ${B}/{A}$, aumenta
rapidamente \`{a} medida que $A$ aumenta e praticamente estabiliza-se em torno
de 8 MeV \cite{almeida:tauhata}. O fato da energia de
liga\c{c}\~ao por n\'{u}cleon n\~ao ser proporcional ao n\'umero de massa $A$
\'e devido ao curto alcance das for\c{c}as nucleares.

Entre 1935 e 1936, Carl Friedrich von Weiszacker e Hans Bethe
aperfei\c{c}oaram uma f\'{o}rmula semi-emp\'{\i}rica, como fun\c{c}\~ao de $A$
e $Z$, para a energia de liga\c{c}\~ao do n\'ucleo at\^omico. Esta f\'ormula,
tamb\'em conhecida como Modelo da Gota L\'{\i}quida \cite{wong}, foi
inicialmente desenvolvidada por George Gamow e Werner Heisemberg. No Modelo da
Gota L\'{\i}quida, o n\'{u}cleo \'e uma esfera cuja densidade \'{e} constante
e a energia de liga\c{c}\~ao do n\'ucleo at\^omico \'e modelada por%

\begin{align}
B(A,Z)  &  = a_{v} A - a_{s} A^{{2}/{3}} - a_{c} \frac{Z(Z-1)}{A^{\frac{1}{3}%
}}\nonumber\\
&  - a_{a}\frac{(A-2Z)^{2}}{A} + \Delta, \label{BWformula}%
\end{align}

\noindent em que cada termo da express\~ao acima representa uma propriedade
nuclear. A seguir discutiremos cada termo da f\'ormula de Bethe-Weizsacker
(\ref{BWformula}): \vskip 0.3cm \noindent$\bullet$ Termo de volume $a_{v}$.

\noindent Como a energia de liga\c{c}\~{a}o por n\'{u}cleon \'{e}
aproximadamente constante, a energia de liga\c{c}\~{a}o deve ser linear com o
volume (proporcionalidade com rela\c{c}\~{a}o ao volume implica a
proporcionalidade com rela\c{c}\~{a}o ao n\'{u}mero de n\'{u}cleons).
Por\'{e}m, o termo de volume considera que todos os n\'{u}cleons est\~{a}o
rodeados por outros n\'{u}cleons, o que n\~{a}o \'{e} verdade para os
n\'{u}cleons pr\'{o}ximos \`{a} superf\'{\i}cie. Assim, o termo de volume
\'{e} superestimado e por isso h\'{a} corre\c{c}\~{o}es representadas pelos
demais termos. \vskip0.3cm \noindent$\bullet$ Termo de superf\'{\i}cie $a_{s}$.

\noindent Os n\'{u}cleons na superf\'{\i}cie do n\'ucleo possuem menos
liga\c{c}\~{o}es com os vizinhos do que os n\'{u}cleons no interior do
n\'{u}cleo, de modo que contribuem menos para a energia de liga\c{c}\~ao
total. O termo de superf\'{\i}cie tem sinal contr\'{a}rio do termo de volume,
pois corrige o valor superestimado deste \'ultimo. \vskip 0.3cm \noindent
$\bullet$ Termo coulombiano $a_{c}$.

\noindent Embora a energia de liga\c{c}\~ao seja devida, principalmente, \`{a}
for\c{c}a nuclear forte, ela \'{e} reduzida devido \`{a} repuls\~ao
Coulombiana entre os pr\'{o}tons. A energia de repuls\~ao Coulombiana \'e
proporcional a $Z^{2}$ e inversamente proporcional ao raio nuclear
$A^{\frac{1}{3}}$. \vskip 0.3cm \noindent$\bullet$ Termo de assimetria
$a_{assim}$.

\noindent Este termo \'e necess\'ario devido a um efeito qu\^{a}ntico
decorrente do Princ\'{\i}pio de Exclus\~ao de Pauli \cite{Piza}.
Experimentalmente verifica-se que os n\'{u}cleos com o n\'{u}mero de
pr\'{o}tons igual ao n\'{u}mero de n\^{e}utrons, $N=Z$, s\~ao mais est\'aveis.
Logo, este termo favorece os n\'ucleos com $N=Z$. Por outro lado, para
n\'ucleos com $N \neq Z$, a energia de liga\c{c}\~ao diminui, devido ao modo
como os n\'{u}cleons se distribuem ao obedecer ao Princ\'{\i}pio de Exclus\~ao
de Pauli. \vskip 0.3cm \noindent$\bullet$ Termo de emparelhamento $\Delta$.

\noindent Este termo captura o efeito do acoplamento de spin (emparelhamento).
Devido novamente ao princ\'{\i}pio de exclus\~{a}o de Pauli, o n\'{u}cleo tem
uma energia de liga\c{c}\~{a}o maior se o n\'{u}mero de pr\'{o}tons com spin
para cima for igual ao n\'{u}mero de pr\'{o}tons com spin para baixo. Isto
tamb\'{e}m \'{e} verdade para os n\^{e}utrons. Assim, se o n\'{u}mero de
pr\'{o}tons $Z$ e o n\'{u}mero de n\^{e}utrons $N$ forem pares, a energia de
liga\c{c}\~{a}o do n\'{u}cleo \'{e} favorecida. Caso contr\'{a}rio, se os
n\'{u}meros $Z$ e $N$ forem \'{\i}mpares, ent\~{a}o a energia de
liga\c{c}\~{a}o \'{e} desfavorecida. Consistentemente com esta an\'{a}lise, na
f\'{o}rmula de massa (\ref{BWformula}) escreve-se que%

\begin{align}
\Delta= \left\{
\begin{array}
[c]{rl}%
+\delta,\;\text{se A e Z forem pares}, & \\
0, \text{se A par e Z \'{\i}mpar(ou vice-versa)}, & \\
-\delta,\;\text{se A e Z forem \'{\i}mpares}. &
\end{array}
\right.  \label{BWformula1}%
\end{align}

\noindent Em \cite{almeida:tauhata} pode-se observar a contribui\c{c}\~ao
de cada termo da f\'ormula de massa de Bethe-Weiszacker
para a energia de liga\c{c}\~ao por n\'ucleon $\frac{B}{A}$, em fun\c{c}\~ao
do n\'{u}mero de massa $A$.

Enfim, no contexto deste trabalho, interessa-nos discutir o termo de
emparelhamento nuclear da f\'{o}rmula de Bethe-Weiszacker (\ref{BWformula}).
Experimentalmente, os n\'{u}cleos com o n\'{u}mero de pr\'{o}tons $Z$ e o
n\'{u}mero de n\^{e}utrons $N$, ambos pares, apresentam uma maior energia de
liga\c{c}\~{a}o por n\'{u}cleon, $B/A$, veja equa\c{c}\~{a}o (\ref{BWformula1}%
). Analogamente ao mecanismo BCS, o aumento da energia de liga\c{c}\~{a}o por
n\'{u}cleon, no caso de n\'{u}cleos par-par, implica a diminui\c{c}\~{a}o da
massa destes n\'{u}cleos, veja equa\c{c}\~{a}o (\ref{massanuclear}). Em outras
palavras, a forma\c{c}\~{a}o de um estado ligado (emparelhado) de dois
n\'{u}cleons, mudou a massa do sistema n\'{u}cleon-n\'{u}cleon. Este mecanismo
\'{e} semelhante ao mecanismo de Higgs, utilizado no Modelo Padr\~{a}o para a
gera\c{c}\~{a}o de massa das part\'{\i}culas elementares. No pr\'{o}ximo
cap\'{\i}tulo, vamos estudar o Modelo Padr\~{a}o e sua lagrangiana, em
particular, o setor do Higgs.

\section{\textbf{Modelo Padr\~ao da F\'{\i}sica Part\'{\i}culas Elementares}}

\label{parteprop}

\noindent A mat\'{e}ria \'{e} formada por mol\'{e}culas e \'{a}tomos que, por
sua vez, s\~ao constitu\'{\i}dos de el\'{e}trons, pr\'{o}tons e n\^{e}utrons.
Sabe-se hoje que pr\'{o}tons e n\^{e}utrons s\~ao formados por quarks
\cite{oldoni.natti:2007}. Nesta se\c{c}\~ao apresentamos primeiramente as
part\'{\i}culas elementares observadas experimentalmente, para em seguida
descrever o Modelo Padr\~ao da f\'{\i}sica de part\'{\i}culas elementares.

\subsection{\textbf{As part\'{\i}culas elementares}}

\noindent Os constituintes b\'{a}sicos da mat\'{e}ria, no sentido de que s\~ao
os blocos de constru\c{c}\~ao da mat\'{e}ria, s\~ao as chamadas
part\'{\i}culas elementares: quarks, l\'{e}ptons e b\'{o}sons mediadores. Para
cada uma destas part\'{\i}culas h\'{a} uma correspondente antipart\'{\i}cula:
part\'{\i}culas com mesma massa, spin e paridade que sua correspondente
part\'{\i}cula, mas com n\'{u}meros qu\^{a}nticos opostos (carga el\'{e}trica,
n\'{u}mero lept\^{o}nico, n\'{u}mero bari\^{o}nico, estranheza, etc.). Na
sequ\^{e}ncia passamos a descrever as part\'{\i}culas elementares observadas
na Natureza.

\subsubsection{\textbf{Quarks}}

\label{quarks}

\noindent Os quarks \cite{oldoni.natti:2007} s\~ao part\'{\i}culas elementares
fermi\^onicas (part\'{\i}culas com spin semi-inteiro) que podem interagir
atrav\'{e}s de todas as quatro intera\c{c}\~oes fundamentais: intera\c{c}\~ao
eletromagn\'{e}tica, forte, fraca e gravitacional. Devido a um fen\^{o}meno
conhecido por confinamento, os quarks n\~ao s\~ao encontrados na Natureza de
forma isolada. Os quarks somente s\~ao observados, indiretamente, em estados
ligados denominados h\'{a}drons.

H\'{a} seis tipos de quarks, tamb\'em denominados sabores, que s\~ao o quark
$u$ (\textit{up}), o quark $d$ (\textit{down}), o quark $s$ (\textit{strange}%
), o quark $c$ (\textit{charm}), o quark $b$ (\textit{bottom}) e o quark $t$
(\textit{top}); e cada quark possui a chamada carga cor: $R$ (\textit{red}),
$B$ (\textit{blue}) e $G$ (\textit{green}). A carga cor dos quarks,
equivalente \`a carga el\'etrica dos el\'etrons, \'e respons\'avel pelo
confinamento dos quarks, pois somente estados (h\'adrons) sem cor podem ser
observados. Assim, de acordo com a Cromodin\^amica Qu\^antica (QCD), somente
estados com tr\^es quarks (um de cada cor), denominados b\'arions, ou estados
quark-antiquark (cor-anticor), denominados m\'esons, podem ser observados experimentalmente.

Os quarks possuem spin $1/2$ e carga el\'{e}trica. Os quarks $u$, $c$ e $t$,
ditos superiores, carregam carga el\'{e}trica $2/3$ e os quarks $d$, $s$ e $b
$, ditos inferiores, carregam carga el\'{e}trica $-1/3$. Os quarks s\~{a}o
classificados em tr\^{e}s fam\'{\i}lias, ou tr\^{e}s gera\c{c}\~{o}es, em que
cada fam\'{\i}lia cont\'{e}m um dubleto, dito de m\~{a}o-esquerda
($L$-\textit{left}), e dois singletos, ditos de m\~{a}o-direita ($R$%
-\textit{right}), ou seja,%

\[
\underbrace{\binom{u_{L}}{d_{L}}, u_{R}, d_{R}}_{\text{Fam\'{\i}lia I}};
\quad\underbrace{\binom{c_{L}}{s_{L}}, c_{R}, s_{R}}_{\text{Fam\'{\i}lia II}};
\quad\underbrace{\binom{t_{L}}{b_{L}}, t_{R}, b_{R}}_{\text{Fam\'{\i}lia III}}%
\]

\subsubsection{\textbf{L\'{e}ptons}}

\label{leptons}

\noindent Os l\'{e}ptons s\~ao part\'{\i}culas elementares fermi\^onicas que
podem interagir atrav\'{e}s das intera\c{c}\~oes eletromagn\'{e}tica, fraca e
gravitacional. Os l\'eptons n\~ao s\~ao sujeitos \`a intera\c{c}\~ao forte e
n\~ao s\~ao constitu\'{\i}dos de quarks. Tamb\'em existem seis tipos, ou
sabores, de l\'{e}ptons, o el\'etron $e$, o m\'uon $\mu$, o tau $\tau$, o
neutrino eletr\^onico ${\nu}_{e}$, o neutrino mu\^onico ${\nu}_{\mu}$ e o
neutrino tau\^onico ${\nu}_{\tau}$. Eles s\~ao classificados em tr\^{e}s
fam\'{\i}lias, em que cada fam\'{\i}lia possui um dubleto de m\~ao-direita e um
singleto de m\~ao-esquerda, ou seja,%

\[
\underbrace{\binom{{{\nu}_{e}}_{L}}{e_{L}}, e_{R}}_{\text{Fam\'{\i}lia I}};
\quad\underbrace{\binom{{{\nu}_{\mu}}_{L}}{{\mu}_{L}}, {\mu}_{R}%
}_{\text{Fam\'{\i}lia II}}; \quad\underbrace{\binom{{{\nu}_{\tau}}_{L}}{{\tau
}_{L}}, {\tau}_{R}}_{\text{Fam\'{\i}lia III}}%
\]

Os tr\^{e}s neutrinos foram e continuam sendo um desafio ao Modelo Padr\~{a}o.
Eles n\~{a}o t\^{e}m cor ou carga el\'{e}trica e assim n\~{a}o interagem via
for\c{c}a forte ou eletromagn\'{e}tica, apenas atrav\'{e}s da for\c{c}a fraca
e gravitacional, se tiverem massa; consequentemente, s\~{a}o dif\'{\i}ceis de
serem observados.

Desde o final dos anos 1960, havia evid\^{e}ncias do d\'{e}ficit de neutrinos
no fluxo de neutrinos solares \cite{Davis:1964}. Muitos experimentos
realizados com detectores \`{a} base de cloro e \'{a}gua-Cerenkov mediram este
d\'{e}ficit. Contudo, somente em 1988, o observat\'{o}rio KamiokaNDE (Kamioka
\textit{Nucleon Decay Experiment}), localizado no monte Kamioka, em Hida,
Jap\~{a}o, confirmou que o n\'{u}mero de neutrinos solares era menor que o
previsto pela teoria, o que estava em desacordo com o Modelo Padr\~{a}o. Esta
comprova\c{c}\~{a}o somente foi poss\'{\i}vel devido \`{a} capacidade do
experimento Kamiokande para observar a dire\c{c}\~{a}o dos el\'{e}trons
produzidos nas intera\c{c}\~{o}es com os neutrinos, o que permitiu determinar
diretamente quais neutrinos eram provenientes do Sol. Posteriormente, o
observat\'{o}rio KamiokaNDE foi melhorado, passando a ser denominado
Super-Kamiokande, o que possibilitou em 1998 a primeira observa\c{c}\~{a}o de
evid\^{e}ncias da oscila\c{c}\~{a}o de neutrinos \cite{Neutrinos:1998}. Esta
foi a primeira observa\c{c}\~{a}o experimental apoiando a teoria de que os
neutrinos t\^{e}m massas diferentes de zero, uma possibilidade que os
f\'{\i}sicos te\'{o}ricos j\'{a} haviam especulado por muitos anos. Era a
evid\^{e}ncia de uma nova f\'{\i}sica \cite{Gusso:2005}.

A resposta para o Problema do D\'{e}ficit de Neutrinos Solares veio somente em
2001, com os resultados experimentais obtidos no Observat\'{o}rio de Neutrinos
SudBury, em Ont\'{a}rio, Canad\'{a}. As observa\c{c}\~{o}es comprovaram a
hip\'{o}tese dos f\'{\i}sicos Gribov e Pontecorvo, de 1969, sobre a
mudan\c{c}a de sabores dos neutrinos, enquanto viajavam do Sol \`{a} Terra, a
denominada Teoria das Oscila\c{c}\~{o}es dos Neutrinos \cite{Gribov:1969}. No
Observat\'{o}rio SudBurry, levando em conta as medidas feitas no
Super-Kamiokande, foi poss\'{\i}vel determinar o n\'{u}mero total de neutrinos
solares e qual a fra\c{c}\~{a}o desse n\'{u}mero correspondia a neutrinos
eletr\^{o}nicos. O n\'{u}mero total de neutrinos eletr\^{o}nicos observados
concordava com as predi\c{c}\~{o}es te\'{o}ricas de Gribov e Pontecorvo
\cite{Gribov:1969}.

Enfim, permanecem as quest\~{o}es: Como as massas dos neutrinos surgem? Por
que as massas dos neutrinos s\~{a}o t\~{a}o pequenas? Por que n\~{a}o s\~{a}o
observados os neutrinos de m\~{a}o direita? No Modelo Padr\~{a}o da
f\'{\i}sica de part\'{\i}culas elementares, os f\'{e}rmions t\^{e}m massa
devido \`{a} intera\c{c}\~{a}o com o campo de Higgs, mas estas
intera\c{c}\~{o}es envolvem ambos f\'{e}rmions, com quiralidade de m\~{a}o
direita e de m\~{a}o esquerda. No entanto, apenas neutrinos de m\~{a}o
esquerda t\^{e}m sido observados at\'{e} agora. Estas e outras quest\~{o}es
sobre os neutrinos permanecem sem respostas!

\subsubsection{\textbf{B\'{o}sons mediadores}}

\label{bosonmediador}

\begin{figure}[h]
\centering
\includegraphics[scale=0.6]{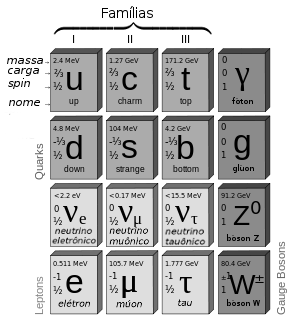}\caption{Representa{c}\~ao das
tr\^{e}s fam\'{\i}lias de f\'{e}rmions em diferentes tonalidades de cinza.}%
\label{esquema:fermions}%
\end{figure}

\noindent Os b\'{o}sons mediadores s\~ao part\'{\i}culas elementares que
possuem spin inteiro e que intermediam as intera\c{c}\~{o}es entre os
f\'{e}rmions. Os b\'{o}sons $W^{+}$, $W^{-}$ e $Z^{0}$ s\~ao mediadores da
intera\c{c}\~ao fraca, os f\'{o}tons ($\gamma$) s\~ao mediadores da
intera\c{c}\~ao eletromagn\'{e}tica e os gl\'{u}ons ($g$) mediam a
intera\c{c}\~ao forte. A figura (\ref{esquema:fermions}) apresenta os
f\'{e}rmions e b\'{o}sons elementares conhecidos.

Na sequ\^{e}ncia descrevemos, por meio dos diagramas de Feynman, como ocorrem
as intera\c{c}\~{o}es entre as part\'{\i}culas elementares.

\subsection{\textbf{As intera\c{c}\~{o}es das part\'{\i}culas elementares}}

\noindent As intera\c{c}\~oes entre as part\'{\i}culas elementares podem ser
representadas diagramaticamente atrav\'{e}s dos chamados diagramas de Feynman.
Este conjunto de regras diagram\'{a}ticas foi introduzido pelo f\'{\i}sico
norte-americano Richard Philips Feymann. O c\'{a}lculo de amplitudes de
probabilidade em F\'{\i}sica de Part\'{\i}culas requer a resolu\c{c}\~ao de
v\'{a}rias integrais muito complicadas, que possuem um grande n\'{u}mero de
vari\'{a}veis. Estas integrais, no entanto, t\^{e}m uma estrutura regular, e
seus resultados podem ser obtidos graficamente atrav\'{e}s de diagramas. Os
diagramas de Feynman s\~ao um conjunto de regras colocadas de forma gr\'afica
para representar as express\~{o}es matem\'{a}ticas que governam as
intera\c{c}\~oes entre as part\'{\i}culas elementares.

Num diagrama de Feynman o eixo vertical representa o tempo e o eixo horizontal
n\~ao representa a dist\^{a}ncia entre as part\'{\i}culas interagentes.
Part\'{\i}culas que caminham no tempo, ditas externas, s\~ao part\'{\i}culas
reais e representam o processo f\'{\i}sico. As part\'{\i}culas internas, que
n\~ao caminham no tempo, s\~ao chamadas de part\'{\i}culas virtuais e n\~ao
s\~ao observadas diretamente; elas representam, nos diagramas de Feynman, os
mecanismos envolvidos nas intera\c{c}\~oes. Para exemplificar, consideremos
diagramas de Feynman da Teoria Eletrofraca que apresenta uma descri\c{c}\~ao
unificada das for\c{c}as eletromagn\'etica e nuclear fraca. Embora estas duas
for\c{c}as manifestem-se de modo muito diferente a baixas energias, a Teoria
Eletrofraca modela-as como dois diferentes aspectos de uma mesma for\c{c}a, a
for\c{c}a eletrofraca, para energias acima da energia de unifica\c{c}\~ao, da
ordem de $10^{2}$ GeV.

\begin{figure}[h]
\centering
\includegraphics[scale=0.3]{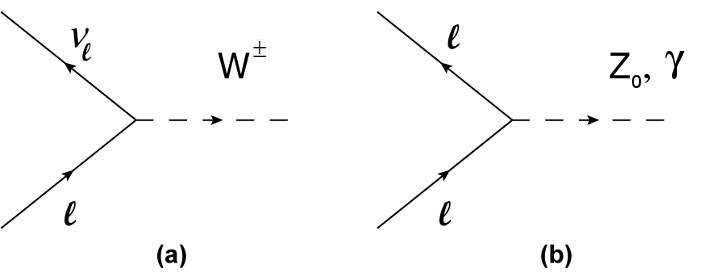}\caption{V\'{e}rtices fundamentais da
Teoria Eletrofraca. (a) V\'{e}rtices com b\'osons carregados. (b) V\'{e}rtices
com b\'osons neutros.}%
\label{diagramaeltrofraco}%
\end{figure}

\begin{figure}[h]
\centering
\includegraphics[scale=0.2]{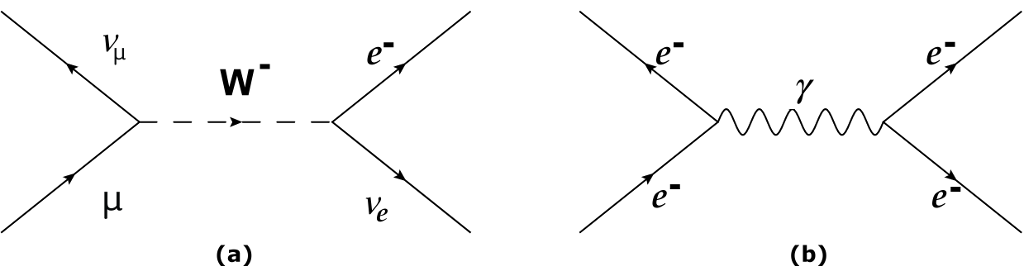}\caption{Alguns processos
eletrofracos. (a) Intera\c{c}\~ao ${\mu}^{-} + {\nu}_{e} \rightarrow{\nu}%
_{\mu} + e^{-}$ atrav\'{e}s do b\'{o}son $W^{-}$. (b) Espalhamento $e^{-} +
e^{-} \rightarrow e^{-} + e^{-}$ atrav\'{e}s do f\'{o}ton $\gamma$.}%
\label{diagramaeltrofraco2}%
\end{figure}

Os diagramas fundamentais de Feynman (v\'ertices) para a intera\c{c}\~ao
eletrofraca s\~ao de dois tipos: os v\'ertices com b\'osons carregados e os
v\'ertices com b\'osons neutros, conforme ilustrados na figura
(\ref{diagramaeltrofraco}). Os b\'{o}sons $W$ carregam carga e transformam os
l\'{e}ptons em seus neutrinos correspondentes, como ilustrado na figura
(\ref{diagramaeltrofraco}a). Por exemplo, o b\'{o}son $W^{-}$ carrega carga
negativa e transforma um el\'{e}tron $e^{-}$ em um neutrino eletr\^{o}nico
${\nu}_{e}$. J\'{a} os b\'osons $\gamma$ e $Z$ n\~ao carregam carga, de modo
que eles explicam, por exemplo, o espalhamento de l\'eptons, como ilustrado na
figura (\ref{diagramaeltrofraco}b). Como exemplo, a figura
(\ref{diagramaeltrofraco2}) apresenta duas poss\'{\i}veis intera\c{c}\~{o}es
em acordo com a Teoria Eletrofraca. O diagrama de Feynman
(\ref{diagramaeltrofraco2}a) representa a intera\c{c}\~ao entre um ${\mu}^{-}$
e um ${\nu}_{e}$ atrav\'{e}s de um $W^{-}$, que transforma o ${\mu}^{-}$ em
${\nu}_{\mu} $ e o ${\nu}_{e}$ em $e^{-}$. Por outro lado, o diagrama de
Feynman da figura (\ref{diagramaeltrofraco2}b) representa o espalhamento de
el\'{e}trons mediado por um f\'{o}ton $\gamma$. Note que em ambos processos
representados na figura (\ref{diagramaeltrofraco2}), a carga \'e conservada.

Na pr\'{o}xima subse\c{c}\~ao apresentaremos o Modelo Padr\~ao da F\'{\i}sica
de Part\'{\i}culas Elementares e seus problemas, dentre eles a necessidade do
B\'{o}son de Higgs e de um mecanismo de gera\c{c}\~ao de massa para as
part\'{\i}culas massivas deste modelo.

\subsection{\textbf{Lagrangiana do Modelo Padr\~ao}}

\label{modpad}

\noindent O Modelo Padr\~{a}o da f\'{\i}sica de part\'{\i}culas elementares
\'{e} um conjunto de teorias matem\'{a}ticas com o objetivo de descrever as
intera\c{c}\~{o}es entre as part\'{\i}culas elementares observadas por meio
das for\c{c}as fraca, forte e eletromagn\'{e}tica. O Modelo Padr\~{a}o se
ajusta com grande precis\~{a}o aos resultados experimentais e \'{e}
consistente com a mec\^{a}nica qu\^{a}ntica e com a relatividade especial,
sendo considerada uma excelente teoria para descrever as intera\c{c}\~{o}es da
natureza. Por\'{e}m, n\~{a}o \'{e} uma teoria completa das intera\c{c}\~{o}es
fundamentais, pois n\~{a}o incorpora a teoria da gravita\c{c}\~{a}o, descrita
pela relatividade geral, e a f\'{\i}sica da mat\'{e}ria ou da energia escura,
al\'{e}m de n\~{a}o descrever alguns fen\^{o}menos, como a oscila\c{c}\~{a}o
dos neutrinos. Os problemas do Modelo Padr\~{a}o e alguns modelos alternativos
ser\~{a}o discutidos na \'{u}ltima se\c{c}\~{a}o desse artigo. Passamos agora
a descrever o Modelo Padr\~{a}o.

A lagrangiana do Modelo Padr\~{a}o pode ser escrita na forma
\cite{Cox-Forshaw:2009}%

\begin{align}
&  \mathcal{L}=-\frac{1}{4}{W}_{\mu\nu}{W}^{\mu\nu}-\frac{1}{4}{B}_{\mu\nu}%
{B}^{\mu\nu}-\frac{1}{4}{G}_{\mu\nu}{G}^{\mu\nu}+\nonumber\label{modelpadr}\\
&  +\overline{\Psi}_{j}{\gamma}^{\mu}\left(  i{\partial}_{\mu}-g{\tau}%
_{j}\cdot{W}_{\mu}-g^{\prime}Y_{j}B_{\mu}-g_{s}\mathbb{T}_{j}\cdot
\mathbb{G}_{\mu}\right)  {\Psi}_{j}+\nonumber\\
&  +|D_{\mu}{\phi}|^{2}+{\mu}^{2}|\phi|^{2}-\lambda|\phi|^{4}-\nonumber\\
&  -\left(  y_{j}\overline{\Psi}_{jL}\phi{\Psi}_{jR}+y_{j}^{\prime}%
\overline{\Psi}_{jL}{\phi}_{c}{\Psi}_{jR}+\mathrm{conjugado}\right)
,\nonumber\\
&
\end{align}
cujos termos v\~{a}o ser explicados nos pr\'{o}ximos par\'{a}grafos.

O Modelo Padr\~{a}o toma como ponto de partida a mat\'{e}ria observada na
natureza. Na figura (\ref{esquema:fermions}), os seis quarks, os tr\^{e}s
l\'{e}ptons carregados e os tr\^{e}s neutrinos associados, al\'{e}m das
antipart\'{\i}culas correspondentes, s\~{a}o representados por $\Psi$. J\'{a}
os quatro b\'{o}sons mediadores da intera\c{c}\~{a}o eletrofraca s\~{a}o
representados por ${W}$ e ${B}$, enquanto os gl\'{u}ons s\~{a}o representados
por ${G}$.

A primeira linha da equa\c{c}\~{a}o (\ref{modelpadr}) representa a energia
cin\'{e}tica carregada pelos b\'{o}sons $W^{+}$, $W^{-}$, $Z$, $\gamma$ e $g$,
dizendo-nos como esses b\'{o}sons interagem uns com os outros.

Na segunda linha da equa\c{c}\~{a}o (\ref{modelpadr}) vemos que as
part\'{\i}culas bos\^{o}nicas (matrizes) ${W}$, ${B}$ e ${G}$ s\~{a}o
envolvidas por part\'{\i}culas fermi\^{o}nicas, matrizes $\Psi$. Estes termos
da lagrangiana do Modelo Padr\~{a}o apontam como os f\'{e}rmions (quarks e
l\'{e}ptons) interagem por meio dos b\'{o}sons mediadores em ${W}$, ${B}$ e
${G}$. As matrizes hermitianas $2\times2$ com tra\c{c}o nulo ${W}$, ${B}$
cont\^{e}m os b\'{o}sons $W^{+}$, $W^{-}$, $Z$ e $\gamma$ , enquanto que as
matrizes hermitianas $3\times3$ com tra\c{c}o nulo ${G}$ cont\^{e}m os
gl\'{u}ons $g$. Enfim, o primeiro termo da segunda linha representa a energia
cin\'{e}tica carregada pelos f\'{e}rmions massivos.

As duas primeiras linhas da lagrangiana do Modelo Padr\~{a}o descrevem o que
ocorre, por exemplo, quando dois el\'{e}trons se aproximam, ou como um certo
tipo de neutrino interage com um anti-muon. Estas intera\c{c}\~{o}es ocorrem
de uma maneira precisa, especificada na lagrangiana (\ref{modelpadr}). Por
exemplo, quando dois quarks interagem atrav\'{e}s da for\c{c}a forte, o
\'{u}nico termo relevante da lagrangiana (\ref{modelpadr}) para descrever esta
intera\c{c}\~{a}o \'{e} o \'{u}ltimo termo da segunda linha, que envolve dois
campos de mat\'{e}ria $\Psi$, interagindo atrav\'{e}s de gl\'{u}ons em ${G}$.
Note que processos (rea\c{c}\~{o}es) que n\~{a}o est\~{a}o presentes na
equa\c{c}\~{a}o (\ref{modelpadr}) n\~{a}o s\~{a}o observados na natureza e vice-versa.

Qual \'{e} a f\'{\i}sica descrita pelas terceira e quarta linhas da
lagrangiana do Modelo Padr\~{a}o? Sabe-se experimentalmente que as
part\'{\i}culas elementares da natureza, relacionadas na figura
(\ref{esquema:fermions}), apresentam simetrias. A lagrangiana do Modelo
Padr\~{a}o (\ref{modelpadr}) incorpora as simetria da natureza. O problema, de
natureza matem\'{a}tica, \'{e} que a natureza exibe uma determinada simetria
de \textit{gauge}, e que todas as part\'{\i}culas do Modelo Padr\~{a}o devem
ser n\~{a}o massivas para que o Modelo Padr\~{a}o apresente esta simetria de
\textit{gauge. }Por outro lado, quarks e l\'{e}ptons carregados s\~{a}o
massivos. Para resolver este problema, Peter Higgs sugeriu um mecanismo
semelhante ao mecanismo encontrado na supercondutividade e no emparelhamento
nuclear. Analogamente aos f\^{o}nons dos pares de Cooper, Higgs sugeriu um
b\'{o}son que daria massa \`{a}s part\'{\i}culas elementares massivas, o
b\'{o}son de Higgs ${\phi}$.

A terceira e a quarta linhas da lagrangiana do Modelo Padr\~{a}o
(\ref{modelpadr}) descrevem a f\'{\i}sica dos b\'{o}sons de Higgs ${\phi}$. A
terceira linha descreve a energia cin\'{e}tica, a massa e a
autointera\c{c}\~{a}o dos b\'{o}sons de Higgs, enquanto a quarta linha
descreve como estes b\'{o}sons interagem com a mat\'{e}ria, f\'{e}rmion $\Psi
$, gerando massa.

Na sequ\^encia descreveremos o Mecanismo de Higgs no Modelo Padr\~ao. Para isto
analisaremos com mais detalhes o setor de Higgs da lagrangiana do Modelo Padr\~{a}o
(\ref{modelpadr}). Adiantamos
que o campo de Higgs \'{e} um campo escalar complexo, mas iremos trabalhar
primeiramente com um campo escalar real com o objetivo de facilitar o processo
de compreens\~{a}o do m\'{e}todo. Considere um campo escalar real $\phi$
descrito pela lagrangiana%

\begin{equation}
\mathcal{L}=\frac{1}{2}({\partial}_{\mu}\phi)^{2}-\underbrace{\left(  \frac
{1}{2}{\mu}^{2}{\phi}^{2}+\frac{1}{4}\lambda{\phi}^{4}\right)  }_{V(\phi)},
\label{lagrhiggs1}%
\end{equation}

\noindent com $\lambda>0$ e $\mu^{2}<0$. Note que a lagrangiana acima e a
terceira linha da lagrangiana (\ref{modelpadr}) t\^em a mesma estrutura.

\begin{figure}[h]
\centering
\includegraphics[scale=0.25]{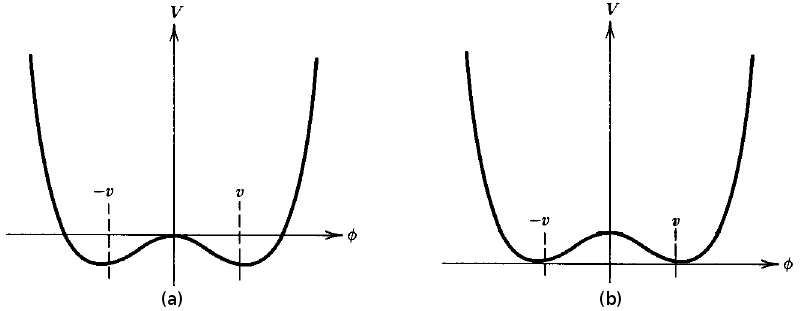}\caption{(a) Potencial $V(\phi) =
\frac{1}{2}{\mu}^{2} {\phi}^{2} + \frac{1}{4}\lambda{\phi}^{4}$ para ${\mu
}^{2} < 0$ e $\lambda> 0$. (b) Potencial deslocado.}%
\label{potmu1}%
\end{figure}

A lagrangiana
(\ref{lagrhiggs1}) \'{e} invariante sob a transforma\c{c}\~{a}o $\phi
\rightarrow-\phi$. O potencial $V(\phi)$ possui a forma dada na figura
(\ref{potmu1}a) com dois pontos de m\'{\i}nimo satisfeitos por%

\begin{equation}
\frac{\partial V}{\partial\phi} = \phi({\mu}^{2} + \lambda{\phi}^{2}) = 0
\quad\Rightarrow\quad{\phi}^{2} = -\frac{{\mu}^{2}}{\lambda} \equiv v^{2}\;.
\end{equation}

\noindent O ponto $\phi=0$ corresponde ao v\'{a}cuo do sistema descrito por
(\ref{lagrhiggs1}) . Nesta configura\c{c}\~{a}o, o v\'{a}cuo n\~{a}o
corresponde ao valor m\'{\i}nimo do campo (sistema). Al\'{e}m disso, a massa
$m$ do campo $\phi$ \'{e} imagin\'{a}ria, o que \'{e} um problema. Para
corrigir estes problemas, redefine-se o campo escalar $\phi$, deslocando
verticalmente seus m\'{\i}nimos para $V=0$, conforme a figura (\ref{potmu1}b).
Assim, define-se um novo campo ${\phi}^{\prime}$ deslocado de ${\phi}$ por
$\eta=\sqrt{-{\mu}^{2}/\lambda}$. O campo $\eta$ pode ser interpretado como
flutua\c{c}\~{o}es qu\^{a}nticas de $\phi$ \cite{Halzem}. Portanto, fazendo
${\phi}^{\prime}={\phi}+\eta$ em (\ref{lagrhiggs1}) obtem-se%

\begin{equation}
\mathcal{L} = \frac{1}{2}({\partial}_{\mu} {\phi}^{\prime})^{2} - \frac{1}{2}
\left(  \sqrt{-2{\mu}^{2}}\right)  ^{2} {\phi^{\prime}}^{2} - \lambda\eta
{\phi^{\prime}}^{3} - \frac{1}{4}\lambda{\phi^{\prime}}^{4} \;.
\label{potmu11}%
\end{equation}

\noindent Na lagrangiana (\ref{potmu11}) o v\'{a}cuo corresponde ao zero do
campo ${\phi^{\prime}}$ e o campo ${\phi^{\prime}}$ possui massa real e
positiva, $m_{{\phi^{\prime}}}=\sqrt{-2{\mu}^{2}}$. Por outro lado, note que a
lagrangiana (\ref{potmu11}) n\~{a}o possui mais a simetria sob a
transforma\c{c}\~{a}o ${\phi^{\prime}}\rightarrow-{\phi^{\prime}}$, devido ao
termo ${\phi^{\prime}}^{3}$. Em analogia aos mecanismos BCS e de
emparelhamento nuclear, dizemos que a quebra de simetria de (\ref{lagrhiggs1})
gerou massa para o campo ${\phi}$, ou seja, $m_{{\phi}}=\sqrt{{\mu}^{2}%
}\rightarrow m_{{\phi^{\prime}}}=\sqrt{-2{\mu}^{2}}$.

No caso da lagrangiana de Higgs, ${\phi}$ \'e um campo escalar complexo da forma%

\begin{align}
\phi= \sqrt{\frac{1}{2}}
\begin{pmatrix}
{\phi}_{1} + i{\phi}_{2}\\
{\phi}_{3} + i{\phi}_{4}%
\end{pmatrix}
\;,
\end{align}

\noindent descrito pela lagrangiana \cite{Halzem}%

\begin{align}
\mathcal{L} = ({\partial}_{\mu} \phi)^{\dagger}({\partial}^{\mu} \phi) - {\mu
}^{2} {\phi}^{\dagger}\phi- \lambda({\phi}^{\dagger}\phi)^{2} \;.
\label{potmu21}%
\end{align}

\indent A lagrangiana (\ref{potmu21}) \'e invariante por uma mudan\c{c}a de
fase local, $\phi\rightarrow e^{i q {\alpha(x)}} \phi$. Pode-se escolher
inicialmente o valor esperado do v\'acuo do campo de Higgs, de modo que
tomando ${\phi}_{1} = {\phi}_{2} = {\phi}_{4} = 0$ e ${\phi}_{3} =
\sqrt{-\frac{{\mu}^{2}}{\lambda}}$, tem-se
\begin{align}
\phi= \sqrt{\frac{1}{2}}
\begin{pmatrix}
0\\
\sqrt{-\frac{{\mu}^{2}}{\lambda}}%
\end{pmatrix}
\;.
\end{align}

\noindent Fazendo um deslocamento semelhante ao descrito acima
\cite{Halzem,Novaes}, envolvendo a derivada covariante em termos dos campos
bos\^onicos $W^{+}$, $W^{-}$, $Z$ e $\gamma$,%

\begin{align}
{\partial}_{\mu} \rightarrow D_{\mu}={\partial}_{\mu} +ig\frac{1}{2}{\tau}
\cdot{W}_{\mu} \;,
\end{align}

\noindent encontra-se um efeito an\'{a}logo ao mostrado para o campo escalar
real. Assim, por meio da quebra de simetria gerada pelo deslocamento do campo
de Higgs, as part\'{\i}culas que a lagrangiana do Modelo padr\~{a}o descreve
adquirem massa. A este formalismo d\'{a}-se o nome de Mecanismo de Higgs. Para
mais detalhes veja as refer\^{e}ncias \cite{Halzem,Novaes}.

\section{\textbf{A procura do Higgs}}

\label{procura}

\noindent No Modelo Padr\~ao a massa do b\'{o}son de Higgs \'{e} dada por
\begin{equation}
m_{H} = \sqrt{\frac{\lambda}{2}}\;v \;,
\end{equation}

\noindent em que $\lambda$ \'{e} o par\^{a}metro de auto-acoplamento do
b\'{o}son de Higgs e $v=246$ GeV \'{e} o valor esperado do campo de Higgs no
v\'{a}cuo, calculado teoricamente \cite{Amsler20081}. O problema \'{e} que o
valor de $\lambda$ \'{e} desconhecido, de modo que o valor da massa do
b\'{o}son de Higgs n\~{a}o pode ser predito teoricamente. A seguir descrevemos
os principais experimentos que procuram observar o b\'{o}son de Higgs.

\subsection{\textbf{FermiLab}}

\begin{figure}[h]
\centering
\includegraphics[scale=0.4]{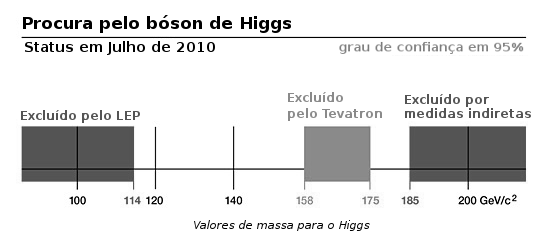}\caption{Espectro permitido para a
massa do Higgs em 2010 \cite{CDF,dzero}.}%
\label{higgs2009}%
\end{figure}

\noindent Experimentalmente, a procura pelo Higgs come\c{c}ou a fornecer
melhores resultados a partir de 2009 por meio de experimentos realizados no
acelerador Tevatron, localizado no Fermilab (\textit{Fermi National
Accelerator Laboratory}), Batavia, Illinois, USA. A an\'{a}lise dos dados da
colabora\c{c}\~{a}o entre os grupos dos detectores CDF (\textit{Collider
Detector at Fermilab}) \cite{CDF} e $D\varnothing$ (\textit{DZero})
\cite{dzero} excluiu algumas fra\c{c}\~{o}es do espectro permitido para a
massa do b\'{o}son de Higgs. Assim, em 2010, o espectro para o qual a massa do
b\'{o}son Higgs era permitida se tornou mais preciso, como mostrado na figura
(\ref{higgs2009}).

Na figura (\ref{higgs2009}) s\~{a}o mostradas as regi\~{o}es de massas do
Higgs que foram exclu\'{\i}das pelos dados experimentais obtidos atrav\'{e}s
dos experimentos CDF e $D\varnothing$, realizados at\'{e} 2010. As \'{a}reas
em branco n\~{a}o foram exploradas devido \`{a} faixa de acelera\c{c}\~{a}o do
Tevatron, de 1 TeV, que \'{e} insuficiente para acessar estas faixas de
energias. O acelerador Tevatron foi desativado em 2011.

\subsection{\textbf{CERN - LHC}}

O LHC (\textit{Large Hadron Collider}) \'{e} atualmente o maior acelerador de
part\'{\i}culas do mundo. O LHC localiza-se em um t\'{u}nel de 27 km de
circunfer\^{e}ncia, a 175 metros abaixo do n\'{\i}vel do solo, na fronteira
franco-su\'{\i}\c{c}a, pr\'{o}ximo a Genebra, na Su\'{\i}\c{c}a. Ele \'{e}
constitu\'{\i}do de quatro grandes detectores e foi projetado para colidir
feixes de part\'{\i}culas com energia de 7 TeV ($7.10^{12}$
el\'{e}tron-volts). O conjunto de detectores de part\'{\i}culas do LHC utiliza
os produtos advindos de uma colis\~{a}o entre dois feixes de part\'{\i}culas
carregadas para reconstruir os processos que ocorreram durante a colis\~{a}o.
A seguir descrevemos os quatro grandes detectores do LHC.

\begin{itemize}
\item[$\bullet$] \textbf{ALICE} (\textit{A Large Ion Collider Experiment}): O
detector ALICE \'{e} especializado para a an\'{a}lise de colis\~{o}es de
\'{\i}ons de chumbo. Tais experimentos visam estudar as propriedades do plasma
quark-gl\'{u}on, um estado da mat\'{e}ria em que quarks e gl\'{u}ons est\~{a}o
sob temperaturas e press\~{o}es muito altas \cite{alicedetetor}.

\item[$\bullet$] \textbf{ATLAS} (\textit{A Toroidal LHC ApparatuS}): O
detector ATLAS \'{e} muito vers\'{a}til, permitindo experimentos que buscam
observar desde o b\'{o}son de Higgs at\'{e} part\'{\i}culas
supersim\'{e}tricas e dimens\~{o}es extras \cite{atlasdetetor}.

\item[$\bullet$] \textbf{CMS} (\textit{Compact Muons Solenoid}): O detector
CMS tem objetivos semelhantes \`aqueles do detector ATLAS \cite{cmsdetetor}.

\item[$\bullet$] \textbf{LHCb} (\textit{Large Hadron Collider beauty}): O
detector LHCb \'{e} especializado no estudo da assimetria entre mat\'{e}ria e
antimat\'{e}ria presente nas intera\c{c}\~{o}es das part\'{\i}culas B
(part\'{\i}culas que cont\^{e}m o quark bottom) \cite{lhcbdetetor}.
\end{itemize}

\subsection{\textbf{A dectec\c{c}\~ao do Higgs}}

Nestes aceleradores, dois feixes de part\'{\i}culas s\~{a}o acelerados em
sentidos contr\'{a}rios, um hor\'{a}rio e outro anti-hor\'{a}rio, pr\'{o}ximos
\`{a} velocidade da luz e, posteriormente, s\~{a}o colimados no interior de um
detector para que colidam e os produtos da colis\~{a}o sejam estudados. Nos
experimentos realizados no LHC, tanto na colabora\c{c}\~{a}o ATLAS como na
colabora\c{c}\~{a}o CMS, dois feixes de pr\'{o}tons colidiram com energia de
$7$ TeV e foram estudados cinco canais de decaimento do b\'{o}son de Higgs
$\phi$, ou seja, $\phi\rightarrow\gamma\gamma$, $\phi\rightarrow b\bar{b}$,
$\phi\rightarrow\tau^{+}\tau^{-}$, $\phi\rightarrow WW$ e $\phi\rightarrow ZZ$.

A produ\c{c}\~{a}o de b\'{o}sons de Higgs em colis\~{o}es
pr\'{o}ton-pr\'{o}ton ocorre por meio de v\'{a}rios canais (rea\c{c}\~{o}es),
cujas \textit{branching ratios} dependem da massa do Higgs. A
\textit{branching ratio} de um decaimento \'{e} a fra\c{c}\~{a}o de
part\'{\i}culas que decaem de um modo espec\'{\i}fico em rela\c{c}\~{a}o ao
n\'{u}mero total de part\'{\i}culas que decaem.

C\'alculos te\'oricos \cite{Amsler20081}
indicam que para massas do Higgs abaixo de $130$
GeV/$c^{2}$, o Higgs ir\'{a} decair principalmente em quarks $b$. Nesta
regi\~ao, o decaimento atrav\'{e}s de pares de taus tamb\'{e}m \'{e}
significativo. J\'a para regi\~{o}es de massas acima de $150$ GeV/$c^{2}$, o
decaimento tem maior probabilidade de ser identificado pelos processos
$H\rightarrow WW$ e $H \rightarrow ZZ$.

A procura de b\'{o}sons de Higgs na regi\~{a}o abaixo de $150$ GeV/$c^{2}$
tamb\'{e}m pode ser realizada atrav\'{e}s do decaimento em pares de
f\'{o}tons \cite{Amsler20081}.
Apesar deste processo possuir uma \textit{branching ratio} muito
pequena, ele fornece uma assinatura clara no calor\'{\i}metro hadr\^{o}nico
(conjunto de medidores para a obten\c{c}\~{a}o da energia de part\'{\i}culas
que interagem fortemente) e proporciona uma excelente resolu\c{c}\~{a}o da
energia no calor\'{\i}metro eletromagn\'{e}tico (conjunto de medidores
utilizado para medir a energia de el\'{e}trons e f\'{o}tons).

At\'e dezembro de 2011, atrav\'{e}s dos dados coletados pelo CMS/LHC (CERN),
Tevatron (FermiLab) e LEP-\textit{Large Electron Positron Collider} (CERN),
observou-se que as regi\~{o}es de
massa, em que o b\'{o}son de Higgs n\~{a}o foi encontrado, com um limite de
confian\c{c}a de 95\%, ficaram restritas \`as massas entre
$114$ GeV/$c^{2}$ a $138$ GeV/$c^{2}$ \cite{Chatrchyan201226}.
Caso ele n\~{a}o
seja observado nesta regi\~{a}o, estar\'{a} comprovado, com limite de
confian\c{c}a de 95\%, que n\~{a}o h\'{a} um b\'{o}son de Higgs como aquele
descrito pelo Modelo Padr\~{a}o. Se esta situa\c{c}\~{a}o for verdadeira, o
Modelo Padr\~{a}o ter\'{a} que ser revisado.

Neste contexto, em 4 de julho de 2012, foi anunciada oficialmente no CERN a
observa\c{c}\~{a}o, nos experimentos ATLAS e CMS, de um sinal que pode ser o
b\'{o}son de Higgs. Resta verificar se a nova part\'{\i}cula observada possui
as propriedades do b\'{o}son de Higgs do Modelo Padr\~{a}o. Se realmente a
nova part\'{\i}cula for identificada como sendo o b\'{o}son de Higgs do Modelo
Padr\~{a}o, ent\~{a}o devemos nos perguntar se o Modelo Padr\~{a}o \'{e}
v\'{a}lido para escalas de energias maiores. Caso contr\'{a}rio, uma nova
f\'{\i}sica estar\'{a} se apresentando para testar os modelos alternativos ao
Modelo Padr\~{a}o.

\section{\textbf{Modelos alternativos}}

\label{modeloalt}

\noindent Neste cap\'{\i}tulo apresentamos os principais problemas do Modelo
Padr\~ao e os principais modelos alternativos ao Modelo Padr\~ao e ao
mecanismo de Higgs.

\subsection{\textbf{Problemas com o Modelo Padr\~{a}o}}

\noindent Como dito anteriormente, o Modelo Padr\~{a}o da f\'{\i}sica de
part\'{\i}culas elementares \'{e} incompleto. Vamos ent\~{a}o mencionar alguns
problemas do Modelo Padr\~{a}o.

\subsubsection{\textbf{Problemas experimentais}}

\begin{itemize}
\item[$\bullet$] \textbf{Intera\c{c}\~{a}o gravitacional}: O Modelo Padr\~{a}o
n\~{a}o fornece uma explica\c{c}\~{a}o da gravidade (qu\^{a}ntica). Al\'{e}m
disto, ele \'{e} incompat\'{\i}vel com a teoria da relatividade geral. Embora
haja a proposta de uma part\'{\i}cula bos\^{o}nica, o gr\'{a}viton, para
explicar a intera\c{c}\~{a}o gravitacional, o Modelo Padr\~{a}o n\~{a}o
descreve esta intera\c{c}\~{a}o.

\item[$\bullet$] \textbf{Oscila\c{c}\~ao de neutrinos}: O Modelo Padr\~ao
n\~ao explica a observa\c{c}\~ao experimental das oscila\c{c}\~oes dos
neutrinos. Uma consequ\^encia desta observa\c{c}\~ao experimental \'e que os
neutrinos teriam massa.

\item[$\bullet$] \textbf{Assimetria mat\'{e}ria - antimat\'{e}ria}: O Modelo
Padr\~{a}o n\~{a}o explica a predomin\^{a}ncia no universo observ\'{a}vel da
mat\'{e}ria sobre a antimat\'{e}ria. Este problema est\'{a} relacionado com a
quebra da simetria CP forte.

\item[$\bullet$] \textbf{Mat\'{e}ria escura - Energia escura}:
Observa\c{c}\~{o}es cosmol\'{o}gicas mostram que o Modelo Padr\~{a}o explica
apenas $4\%$ da mat\'{e}ria ou energia observada no universo.

\item[$\bullet$] \textbf{Infla\c{c}\~{a}o c\'{o}smica}: Observa\c{c}\~{o}es
cosmol\'{o}gicas mostram que o universo sofreu, em seu momento inicial, um
crescimento exponencial, propelido por uma esp\'{e}cie de for\c{c}a
gravitacional repulsiva, ou press\~{a}o negativa. O Modelo Padr\~{a}o n\~{a}o
explica estas observa\c{c}\~{o}es cosmol\'{o}gicas.
\end{itemize}

\subsubsection{\textbf{Problemas Te\'{o}ricos}}

\begin{itemize}
\item[$\bullet$] \textbf{Origem da massa}: No Modelo Padr\~{a}o, a origem das
massas das part\'{\i}culas elementares \'{e} explicada por meio do campo de
Higgs. Observa\c{c}\~{o}es experimentais tentam comprovar a exist\^{e}ncia do
b\'{o}son de Higgs do Modelo Padr\~{a}o.

\item[$\bullet$] \textbf{Excesso de par\^{a}metros}: O Modelo Padr\~{a}o
cont\'{e}m 19 par\^{a}metros livres, como as massas da part\'{\i}culas
elementares, que s\~{a}o determinados experimentalmente e n\~{a}o podem ser
calculados independentemente.

\item[$\bullet$] \textbf{Problema da hierarquia nas intera\c{c}\~{o}es}: O
Modelo Padr\~{a}o n\~{a}o explica porque as intensidades das for\c{c}as
fundamentais s\~{a}o t\~{a}o diferentes. Ele n\~{a}o explica o porqu\^{e} da
for\c{c}a fraca ser $10^{32}$ vezes mais forte que a for\c{c}a gravitacional.

\item[$\bullet$] \textbf{Problema da hierarquia nas fam\'{\i}lias de
l\'{e}ptons}: Toda mat\'{e}ria que constitui os compostos na Terra \'{e}
formada pelos f\'{e}rmions das primeiras fam\'{\i}lias (${\nu}_{e},e^{-}%
,u,d$). As segundas e terceiras fam\'{\i}lias de f\'{e}rmions s\~{a}o
c\'{o}pias muito mais pesadas das primeiras fam\'{\i}lias. O Modelo Padr\~{a}o
n\~{a}o nos d\'{a} uma explica\c{c}\~{a}o para a grande diferen\c{c}a entre as
massas destas fam\'{\i}lias.

\item[$\bullet$] \textbf{Problema da (n\~{a}o) viola\c{c}\~{a}o da simetria CP
forte}: A simetria de conjuga\c{c}\~{a}o de carga--paridade na
intera\c{c}\~{a}o fraca \'{e} violada. Por outro lado, esta simetria na
intera\c{c}\~{a}o forte n\~{a}o \'{e} violada. O Modelo Padr\~{a}o n\~{a}o
explica porque a intera\c{c}\~{a}o nuclear forte \'{e} CP--invariante.
\end{itemize}

Devido a problemas como esses, t\^{e}m sido propostas extens\~{o}es do Modelo
Padr\~{a}o e modelos alternativos. Alguns desses modelos s\~{a}o apresentados
nas pr\'{o}ximas subse\c{c}\~{o}es.

\subsection{\textbf{Modelo supersim\'{e}trico}}

\noindent A supersimetria \'{e} uma proposta de simetria que considera a
exist\^{e}ncia de um parceiro (supersim\'{e}trico) para cada part\'{\i}cula
elementar existente \cite{Kane}. Para cada b\'{o}son (f\'{e}rmion) do Modelo
Padr\~{a}o deve existir um parceiro fermi\^{o}nico (bos\^{o}nico), com massa e
constante de acoplamento id\^{e}nticas. Esta situa\c{c}\~{a}o
supersim\'{e}trica n\~{a}o \'{e} observada na natureza. A id\'{e}ia \'{e} que
a supersimetria tenha ocorrido no universo em seus instantes iniciais e que,
com o resfriamento, esta simetria entre b\'{o}sons e f\'{e}rmions tenha sido
quebrada, como no modelo BCS. Assim, na situa\c{c}\~{a}o atual do universo, as
massas dos parceiros supersim\'{e}tricos est\~{a}o em uma escala de energia
acima da escala acess\'{\i}vel aos aceleradores atuais. A id\'{e}ia do
parceiro supersim\'{e}trico \'{e} atrativa, pois oferece uma solu\c{c}\~{a}o
ao \textquotedblleft problema da hierarquia\textquotedblright.

\begin{figure}[h]
\centering
\includegraphics[scale=0.4]{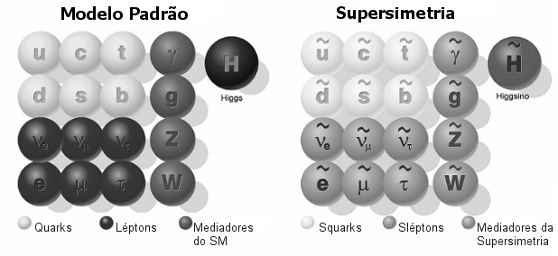}\caption{Part\'{\i}culas e seus
parceiros supersim\'{e}tricos no Modelo Supersim\'{e}trico M\'{\i}nimo (MSSM).}%
\label{sparticulas}%
\end{figure}

A supersimetria tamb\'{e}m oferece uma poss\'{\i}vel solu\c{c}\~{a}o para o
problema da mat\'{e}ria escura. No Modelo Supersim\'{e}trico M\'{\i}nimo
(MSSM), designa-se a part\'{\i}cula elementar supersim\'{e}trica mais leve
(est\'{a}vel) como sendo a LSP (\textit{Lighest Supersimmetric Particle}). Uma
das poss\'{\i}veis candidatas \`{a} LSP \'{e} o neutralino
\cite{JamesPinfold:2005}. Acredita-se que o neutralino tenha sido produzido
nos prim\'{o}rdios do Universo e que ainda exista atualmente devido \`{a} sua
estabilidade, correspondendo \`{a} mat\'{e}ria escura observada. O espectro de
part\'{\i}culas do MSSM \cite{ref-fig14} \'{e} mostrado na figura
(\ref{sparticulas}).

Por outro lado, os v\'arios experimentos conduzidos nos aceleradores de
part\'{\i}culas, que procuraram observar a supersimetria, forneceram medidas
compat\'{\i}veis com as previs\~oes do Modelo Padr\~ao das part\'{\i}culas elementares.
Recentemente, experimentos conduzidos no LHCb e no CMS analisaram os resultados da coleta
de dados do decaimento de m\'esons $B_s$ (m\'esons compostos de um anti-quark bottom
e um quark strange). Esperava-se observar a supersimetria nesses dados,
o que n\~ao ocorreu. At\'e o momento n\~ao h\'a nenhuma confirma\c{c}\~{a}o
experimental da supersimetria.

\subsection{\textbf{Modelo Technicolor}}

\noindent Este modelo foi proposto inicialmente para tentar contornar o
problema da hierarquia e da origem da massa. Ao inv\'{e}s de introduzir o
b\'{o}son (mecanismo) de Higgs para explicar a massa dos b\'{o}sons $W$ e $Z$
por meio de uma quebra espont\^{a}nea de simetria, o Modelo Technicolor gera
massas por meio de um processo din\^{a}mico, envolvendo a intera\c{c}\~{a}o
forte. A id\'{e}ia b\'{a}sica do Modelo Technicolor, introduzida primeiramente
por Steven Weinberg \cite{Weinberg.Tech:1979} e Leonard Susskind
\cite{Susskind.Tech:1979}, em 1979, \'{e} a de que o b\'{o}son de Higgs n\~ao
\'e uma part\'{\i}cula fundamental, mas uma part\'{\i}cula composta de outras
part\'{\i}culas, chamadas techniquarks. Em analogia com a teoria dos quarks,
os techniquarks s\~{a}o mantidos atrav\'{e}s de technigl\'{u}ons. De acordo
com o Modelo Technicolor, a forma como estas technipart\'{\i}culas se agrupam
acaba gerando part\'{\i}culas com diferentes massas, explicando o problema da
hierarquia de massas e a origem da massa.

\subsection{\textbf{Dimens\~{o}es Extras}}

\noindent Outra alternativa para evitar o problema da hierarquia foi sugerida,
em 1998, por Nima Arkani-Hamed, Savas Dimopoulos e Gia Dvali
\cite{Dvali.Dimopoulos.Hamed:1998}. A nova id\'{e}ia \'{e} que, se
considerarmos mais dimens\~{o}es espaciais na Natureza, o potencial
gravitacional emitido por gr\'{a}vitons, observado nas tr\^{e}s dimens\~{o}es
usuais, pode ser menor que o potencial gravitacional total, devido ao fato de
que parte dos gr\'{a}vitons estariam se propagando nas dimens\~{o}es extras.

Nos Modelos de Dimens\~{o}es Extras consideram-se universos $N$-dimensionais,
com $N=4+n$, em que $n$ \'{e} o n\'{u}mero de dimens\~{o}es extras e $4$
corresponde \`{a}s tr\^{e}s dimens\~{o}es espaciais usuais mais a temporal.
Nestes modelos as dimens\~{o}es extras estariam compactadas em dimens\~{o}es
da ordem do comprimento de Planck, $10^{-33}cm$, de modo que seriam
inacess\'{\i}veis \`{a}s nossas observa\c{c}\~{o}es. Enfim, o gr\'{a}viton,
que \'{e} o intermedi\'{a}rio da for\c{c}a gravitacional, interagiria n\~{a}o
somente nas nossas tr\^{e}s dimens\~{o}es espaciais conhecidas, mas tamb\'{e}m
nas outras $n$ dimens\~{o}es. Isso explicaria porque, em nosso espa\c{c}o
tridimensional, a intera\c{c}\~{a}o gravitacional \'{e} t\~{a}o fraca quando
comparada com as outras intera\c{c}\~{o}es.

\section{\textbf{Conclus\~oes}}

\label{conclusao}

\noindent Os pr\'{o}ximos meses ou anos ir\~{a}o responder se o b\'{o}son
observado em 4 de julho de 2012 \'{e} o b\'{o}son de Higgs do Modelo
Padr\~{a}o ou se ele tem outras propriedades, o que indicaria a exist\^{e}ncia
de um Higgs descrito por outro modelo. Note que os v\'{a}rios canais de
decaimento do Higgs, em fun\c{c}\~{a}o da massa do Higgs,
variam de modelo para modelo. As \textit{branching
ratios} do decaimento do Higgs no Modelo Padr\~{a}o, nos Modelos
Supersim\'{e}tricos, nos Modelos Technicolors e nos Modelos de Dimens\~{o}es
Extras s\~{a}o diferentes. Comparando com os resultados esperados a partir do
Modelo Padr\~{a}o, os dados experimentais existentes indicam um excesso de
decaimentos de Higgs em pares de f\'{o}tons, canal $\phi\rightarrow
\gamma\gamma$, e uma car\^{e}ncia nos decaimentos em part\'{\i}culas taus,
canal $\phi\rightarrow\tau^{+}\tau^{-}$. Enfim, o interessante \'{e} o que
est\'{a} por vir. Sabemos que a part\'{\i}cula observada \'{e} um b\'{o}son,
mas n\~{a}o sabemos se corresponde ao b\'{o}son de Higgs do Modelo Padr\~{a}o
ou se \'{e} algo mais ex\'{o}tico. Aguardemos!

\subsection{\textbf{Agradecimentos}}

\indent O autor J. J. M. Pimenta agradece ao CNPq pela bolsa de
Inicia\c{c}\~{a}o Cient\'{\i}fica PIBIC/UEL/CNPq.

\renewcommand\refname{Refer\^{e}ncias}

\end{document}